# Privacy in the Internet of Things: Threats and Challenges

Jan Henrik Ziegeldorf[1*], Oscar Garcia Morchon[2], and Klaus Wehrle[1]

[1] Communication and Distributed Systems, RWTH Aachen University, Aachen, Germany
[2] Philips Research, Eindhoven, Netherlands

## ABSTRACT

The Internet of Things paradigm envisions the pervasive interconnection and cooperation of smart things over the current and future Internet infrastructure. The Internet of Things is, thus, the *evolution* of the Internet to cover the real-world, enabling many new services that will improve people's everyday lives, spawn new businesses and make buildings, cities and transport smarter. Smart things allow indeed for ubiquitous data collection or tracking, but these useful features are also examples of privacy threats that are already now limiting the success of the Internet of Things vision when not implemented correctly. These threats involve new challenges such as the pervasive privacy-aware management of personal data or methods to control or avoid ubiquitous tracking and profiling. This paper analyzes the privacy issues in the Internet of Things in detail. To this end, we first discuss the evolving features and trends in the Internet of Things with the goal of scrutinizing their privacy implications. Second, we classify and examine privacy threats in this new setting, pointing out the challenges that need to be overcome to ensure that the Internet of Things becomes a reality.

## 1. INTRODUCTION

The Internet of Things (IoT) foresees the interconnection of billions to trillions [1, 2], of smart things around us – uniquely identifiable and addressable everyday things with the ability to collect, store, process and communicate information about themselves and their physical environment [3]. IoT systems will deliver advanced services of a whole new kind based on increasingly fine-grained data acquisition in an environment densely populated with smart things. Examples of such IoT systems are pervasive healthcare, advanced building management systems, smart city services, public surveillance and data acquisition, or participatory sensing applications [4, 5].

The increasingly invisible, dense and pervasive collection, processing and dissemination of data in the midst of people's private lives gives rise to serious privacy concerns. Ignorance of those issues can have undesired consequences, e.g. non-acceptance and failure of new services, damage to reputation, or costly law suits. The public boycott of the Italian retailer Benetton in 2003 [6, 7], the revocation of the Dutch smart metering bill in 2009 [8], or the recent outcry against the EU FP7 research project *INDECT* [9, 10] are only three examples of IoT related projects that experienced huge problems due to unresolved privacy issues.

Privacy has been a hot research topic in different technology and application areas that are important enablers of the IoT vision, e.g. RFID, wireless sensor networks (WSN), web personalization, and mobile applications and platforms. Despite considerable contributions from these communities a holistic view of arising privacy issues in the IoT is missing, since the IoT is an *evolving concept* that comprises a growing number of technologies and exhibits a range of changing features. Among these, we witness an explosion in the number of smart things and new ways of interacting with systems and presenting feedback to users. As we will show, these new features of the IoT will aggravate privacy issues and introduce unforeseen threats that pose challenging technical problems. These privacy threats, whether known or new, need to be considered (i) in a reference model of the IoT accounting that accounts for its specific entities and data flows, (ii) from the perspective of existing privacy legislation, and (iii) with regard to the unique and evolving features in the IoT. For without a clear understanding of the arising issues and the appropriate counter-measures, the success of new pioneering services and their users' privacy will be at peril.

We contribute to this discussion in three steps: First, Section 2 frames the notion of privacy in the IoT by discussing its definition, a privacy-aware reference model for the IoT, as well as existing privacy legislation. Second, we contribute a detailed analysis of the IoT evolution from the point of view of the involved technologies and features in Section 3. Third, Section 4 analyzes in detail privacy threats and challenges in the context of our reference model. Section 5 concludes this paper.

## 2. PRIVACY DEFINITION AND REFERENCE MODEL FOR THE IOT

This section provides a framework to our analysis of privacy threats and challenges. Section 2.1 defines privacy in the specific context of the IoT. We observe that most reference models are inadequate for privacy discussions and present our own IoT reference model in Section 2.2. Finally, Section 2.3 discusses the scope of current privacy legislation and its limitations in the IoT context.

### 2.1. Privacy Definition

Privacy is a very broad and diverse notion for which literature offers many definitions and perspectives [11]. From a historic view, the notion of privacy shifted between media, territorial, communication, and bodily privacy. With the increasing use and efficiency of electronic data processing *information privacy* has become the predominant issue today. Information privacy was defined by Westin in 1968 as "the right to select what personal information about me is known to what people" [12]. While Westin's definition, although it referred to non-electronic environments, is still valid, it is also too general to enable focussed discussion about privacy in the IoT. We thus adapt and concretize definition:

*Privacy in the Internet of Things is the threefold guarantee to the subject for*

- *awareness of privacy risks imposed by smart things and services surrounding the data subject*

- *individual control over the collection and processing of personal information by the surrounding smart things*

- *awareness and control of subsequent use and dissemination of personal information by those entities to any entity outside the subject's personal control sphere*

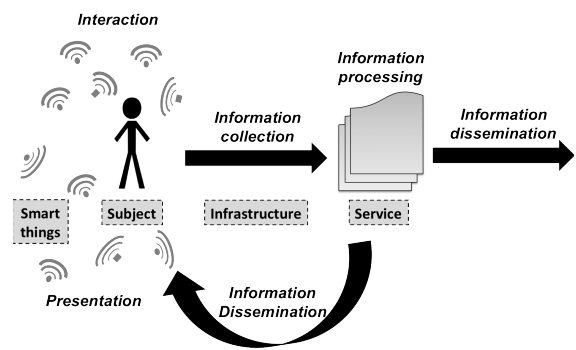

**Figure 1.** IoT reference model with relevant entities and data flows in a typical IoT application.

Our definition of privacy captures in essence the idea of *informational self-determination* by enabling the subject (i) to assess his personal privacy risks, (ii) to take appropriate action to protect his privacy, and (iii) to be assured that it is enforced beyond his immediate control sphere.

The operating systems analogy described by Radomirovic in [13] is a similar concept to characterize what we refer to as the personal sphere of the data subject. In smart home scenarios it can be pictured as that person's household or immediate vicinity, as Radomirovic fittingly observes. However, the exact scope of the subject's personal sphere can differ from situation to situation and it is still unclear what constitutes the individual's personal sphere, or operating system boundaries in the analogous terms, in e.g. a workplace environment or public space.

Similarly, the notion of personal information is necessarily fuzzy, since privacy is a deeply social concept and subject to greatly varying individual perception and requirements [14, 15]. Hence, care must be taken when designing new systems and services to carefully assess the sensitivity of the involved information and relating user requirements, e.g. as businesses are starting to implement in privacy impact analysis's (PIAs). Ultimately, our definition must be understood such that the user may define what he considers personal information.

### 2.2. Reference model

The IoT reference model we propose is based on the ITU [16] and IERC [17] visions of the IoT and can be summarized as: *Anyone and anything is interconnected anywhere at any time via any network participating in any service*. Our reference model describes the entities and information flows of IoT applications.

In the model we consider **four main types of entities** as depicted in Figure 1. *Smart things* are everyday things augmented with information and communication technology (ICT). They are able to collect, process and communicate data about themselves and their environment as well as to interact with other things and humans.

Backends host *services* that gather, combine and analyze data from many smart things to offer a value-added service to the end-user. Humans have two different roles in our reference model. They can be *subject* to data collection by the smart things surrounding them or *recipients* of data or services. Note, that a person can be both subject and recipient at the same time, e.g. in a personal healthcare application. Finally, smart things are connected to services via an *infrastructure* with different characteristics ranging from low-power lossy networks to powerful Internet backbones possibly traversing different intermediate gateways and servers, e.g. firewalls and protocol bridges.

Furthermore, we consider **five different types of information flows**, also depicted in Figure 1, corresponding to the phases in which the *subject* engages in a typical IoT application. In the *Interaction* phase, the data subject actively or passively interacts with the smart things in his environment, thereby triggering a service. Smart things then engage in the *Collection* of information and relay it to the corresponding back-end via the available interconnection networks possibly with the help of intermediate gateways. In the *Processing* phase, backends analyze the information in order to provide the triggered service. *Dissemination* of information towards the data subject and potentially towards third parties constitutes the fourth phase. Finally, in the *Presentation* phase, the service is provided to the data subject by the surrounding smart things according to the instructions by the backend. Note that depending on the location and implementation of the backend services, the flow of information is thus either (i) vertical, in the case of a central distant backend, (ii) horizontal, given a service that runs locally distributed across the smart things, or (iii) hybrid, when one or more local things provide the service.

Our model abstracts from specific device classes, technologies, interconnection methods and services in order to provide a high-level model that fits a wide range of IoT systems and applications. Its assumptions for ubiquitous interconnection of smart things and their capabilities are general and powerful enough to account for the ongoing evolution of the IoT. These assumptions might not yet hold today, but represent the expected developments of the IoT, as shows our analysis (Section 3.2) and are also shared by others, e.g. in Radomirovic's *dense Internet of Things model* [13]. We thus believe that the presented reference model adequately addresses our considerations and requirements.

Note that different reference models have been proposed to account for new technologies and applications as they are added to the wide umbrella that is the IoT today. Among others, the IoT-i consortium [18] and Atzori et. al [5] survey existing reference models and IoT architectures. Considerable progress towards an explicit reference model has been made e.g. by EU FP 7 projects IoT-A [19] and CASAGRAS [20]. However, the proposed models often seem too complex for our purpose or do not fully address the following requirements. Privacy is an exclusively human concern and we need to explicitly consider the roles and involvement of users in the reference model. Furthermore, the model should allow to reason about privacy at a high level based on generic capabilities of things and services. Finally, the IoT is an evolutionary process and the reference model must thus abstract from the underlying things and technologies. Especially, it should not be limited to single technology spaces as e.g. RFID technology.

## 2.3. Privacy legislation

Privacy legislation tries to draw boundaries to the evermore data-hungry business models of many Internet enterprises (e.g data market places, advertising networks, e-commerce sites) and to define mandatory practices and processes for privacy protection. We shortly review the development and practical impact of privacy legislation then identify relevant points and problems in the context of the Internet of Things.

Privacy is recognized as a fundamental human right in the 1948 Universal Declaration of Human Rights and is anchored in the constitutional law of most countries today. The first major piece of legislation on information privacy was passed with the 1974 US Privacy Act, which established the *fair information practices* (FIPs). The FIPs comprise the principles of notice, consent, individual access and control, data minimization, purposeful use, adequate security and accountability. They have been taken up in [21] by the Organization for Economic Co-operation and Development (OECD), which anticipated trade-barriers from the increasingly diverse national privacy legislation. While US privacy legislation continued with a miscellany of specific sectorial laws, the European Union's aim for comprehensive legislation resulted in the 1995 *Directive 95/46/EC on the protection of individuals with regard to processing of personal data and on the free movement of such data* [22]. The directive embeds the FIPs and adds the principle of *explicit consent*, which basically forbids any kind of data collection without explicit permission from the subject.

The practical impact has been different: While the Privacy Act of 1974 was unsuccessful in the US (as were many of its follow-ups), the EU data directive has not only effectively increased data protection standards in Europe, but also evoked international self-regulatory efforts, e.g. the *Safe Harbor* agreement [23]. Today, primarily the principles of notice, consent, access and security are enforced e.g. in e-commerce and online advertising. Privacy legislation also touches some mature technologies which are part of the IoT evolution: RFID and camera networks have received much attention in the past. Recent legislation efforts have focussed on data protection in cloud computing and adequate protection of web users against tracking.

However, already today the level of privacy protection offered by legislation is insufficient, as day-to-day data spills and unpunished privacy breaches [24] clearly indicate. The Internet of Things will undoubtedly create new grey areas with ample space to circumvent legislative boundaries.

First, most pieces of legislation center around the fuzzy notion of *Personally Identifiable Information* (PII). However, efforts towards a concise definition of what constitutes PII (e.g. by enumerating combinations of identifying attributes) are quickly deprecated as new IoT technologies unlock and combine new sets of data that can enable identification and make it increasingly difficult to distinguish PII from non-PII. Our definition of IoT privacy (Section 2.1) acknowledges this problem and demands for involvement of the data subject.

Second, *timeliness* of legislation is a constant issue: E.g. tracking of web-users has been used for many years, before the European Commission passed a law against it in early 2011. With the IoT evolving fast, legislation is bound to fall even farther behind. An example are Smart Meter readings, which already allow to infer comprehensive information about people's lifestyle.

Third, already today, many privacy breaches go unnoticed. In the IoT, *awareness* of privacy breaches among users will be even lower, as data collection moves into everyday things and happens more passively (Table I). Legislation, however, is often only a response to public protests and outcries that require awareness of incidents in the first place.

Finally, the *economics* of privacy are still in favor of those in disregard of privacy legislation. On the one side, development of PETs, enforcement and audits of privacy-protection policies is expensive and can limit business models. On the other side, violations of privacy legislations either go unpunished or result only in comparably small fines, while public awareness is still too low to induce intolerable damage of public reputation. Thus, disregard of privacy legislation, as e.g. Google deliberately circumventing Safari's user tracking protection [25], seems profitable. Over this incident, Google paid a record fine of $22.5 Million in a settlement with the Federal Trade Commission (FTC), but it is conceivable that the earnings more than compensated.

It will be a major challenge to design a unified enduring legislative framework for privacy protection in the Internet of Things, instead of passing quickly outdated pieces of legislation on singular technologies. Success will undoubtedly require a comprehensive knowledge of the technologic basis of the IoT and its ongoing evolution. The key, however, will be a deep understanding of existing and lingering new threats to privacy in the IoT – these threats are what legislation needs to protect against, ultimately.

## 3. EVOLUTION OF THE IOT

Starting with the vision of ubiquitous computing [26], the Internet of Things depends on technologic progress to bring increasing miniaturization and availability of information and communication technology at decreasing cost and energy-consumption. The IoT is thus not a disruptive new technology, but a novel paradigm whose full realization will be a gradual process. This section reviews key technologies in the past evolution of the IoT and related work in privacy (Section 3.1) and then analyzes the current features of the IoT and how they evolve (3.2). This discussion of evolving technologies and features allows us to recognize and analyze privacy threats and challenges early-on.

### 3.1. Evolving Technologies

*RFID technology* stands at the beginning of the IoT vision: it enables passive automatic identification of things at the price of a couple of cents. Indeed, the realization of the IoT vision is still often seen in the pervasive deployment of RFID tags. RFID privacy issues have been thoroughly researched [27, 28, 29]. The dominant threats are automatic *identification* and *tracking* of people through hidden tags, e.g. in clothes. Different countermeasures have been proposed, such as reader authentication, tag encryption, randomizing tag identifiers, and blocking or killing of tags. Managing the lifecycle of an RFID tag is a challenge that will also be interesting for smart things in the IoT context.

*Wireless sensor network* (WSN) technology forms the next evolutionary step of the IoT: Things become active, as they are augmented with sensing, processing and communication capabilities to build first interconnected networks of things. Today, sensor nodes range from tiny millimeter-sized sensor nodes (e.g. Smart Dust) to meter-scale GSM-equipped weather stations. Sensor networks include both small-scale home deployments and also large-scale industrial monitoring systems enabled by standards such as ZigBEE [30], Z-Wave [31], ANT [32] or Bluetooth [33]. Privacy research in WSNs has focussed on privacy threats with regard to the collected sensor data [34, 35], queries to the network [36, 37] and the location of data sources and base stations [38, 39, 40]. Particular challenges that have been commonly identified are uncontrollable environments and resource constraints, which also characteristics of the IoT.

The advent of *smart phones* has further progressed the realization of the IoT vision, being the first mobile mass devices with ubiquitous Internet connection. Smart phones gather critical amounts of private data about their owner, e.g. identifiers, physical location, and activity that bear considerable privacy risks [51]. Extensive privacy research has been conducted focussing mostly on location-based services [52, 53, 54], detection of privacy breaches [55, 56] and privacy aware architectures for participatory sensing [57, 58].

The *cloud computing* paradigm has thrived over the last decade and offers means to handle the expected information explosion in the IoT. Privacy research in cloud computing has focussed on adequate data protection and prevention of information leaks [59, 60], auditing and provenance [61], and private information processing [62, 63]. Cloud platforms are increasingly used to implement backends for storing, processing and accessing information of IoT applications, as e.g. platforms by COSM [64] and Arrayent [65].

**Table I.** Evolving features of the IoT today, in the past and future. Predictions are based on extensive literature research and analysis of past and current developments in the IoT and in related areas. Error margins of predictions are indicated qualitatively.

|  | Technology | Size | Interconnection | Data collection | Thing Interaction | System Interaction | Lifecycle | Vertical vs. horizontal |
|---|---|---|---|---|---|---|---|---|
| **2000** | RFID | Millions | Wired, stationary | Identifier | None | None | Ownership transfer | None |
| **2013** | Sensors, phones, cloud | Billions | Wireless, mobile, H2M | Sensory, limited areas, active humans | Buttons, touch, displays | Smartphone, gestures, speech, web interfaces | Ownership transfer | Mainly vertical |
| **2020** | ICT inside things, new technologies | Billions to trillions | E2E, All-IP, M2M, interoperability | Increasing coverage, passive humans | Haptic, web interfaces | Haptic, using the environment | Product history log, exchangeable | Both horizontal and vertical |
| **Uncertainty** | Invisibility, ubiquity | Billions to trillions | Ubiquity, standards | Extent, penetration | Prevalence of web interfaces | Using all human senses | Dynamicness | Central solutions prevail |
| **References** | [3, 5, 41, 42] | [1, 2, 43, 44, 45, 46] | [2, 3, 41, 46] | [3, 13, 41, 46] | [2, 3, 42, 47] | [2, 3, 47] | [3, 48, 49, 50] | [41, 50] |

### 3.2. Evolving Features

Table I provides a summary of selected features that we consider are most important in the context of privacy. Further and more general discussions of developments in the IoT can be found in [3, 5, 41]. We form qualitative predictions for the future evolution of the selected features based on three pillars: (i) extrapolation from their past and current development, (ii) comparison to trends observed in related areas, and (iii) survey of opinions and predictions from related literature (sample references are provided in the table). Since it is very hard to back such predictions with hard data, errors are necessarily involved. Table I thus also indicates qualitatively the estimated error margin of the predictions.

*Technology:* As described in Section 3.1, the Internet of Things started with the Auto-ID Labs [66] envisioning automatic identification of things through RFID technology. Today, we see that more and more technologies are integrated under the IoT umbrella term, e.g. sensor networks, smart phones, and cloud services. Predictions for the future primarily foresee that information and communication technology increasingly moves into things thereby making them smarter and self-aware [3, 5, 42]. At the same time, new technologies will be invented and made suitable for the masses. Against this background, it is however unclear if, when, and to which extent the envisioned state of ubiquity and invisibility will be reached [41].

*Size:* Estimates from academia and major industry players, among them IBM, HP, and Cisco, predict staggering numbers of 50 billion [1, 3, 41] to trillions of connected things by 2020 [2, 44, 45, 43]. Although predicted numbers and time horizons vary considerably, even defensive scenarios indicate that the IoT will increase the size of networks by orders of magnitude. Such massive amounts of devices will seriously challenge the scalability of existing and forthcoming privacy technology.

*Interconnection:* With the technologic advances and decreasing costs of wireless communications, the IoT is predicted to evolve towards a state where smart things are ubiquitously interconnected [2]. All-IP end-to-end connectivity is the prevalent vision for realization of such pervasive interconnectedness [3, 41, 5] and is heavily pushed by the IETF 6LoWPAN and ETSI M2M working groups. As smart things become query-able from any distance in this process, this will not only bring forth great opportunities for new services but also challenging privacy issues.

*Data collection:* With people carrying smart phones everywhere and sharing their lives on social networks, we are already today witnessing an increasing penetration of people's private and public lives by technology that enables data collection and with it identification, tracking and profiling [13]. With the predicted advent and increasing density of smart things, data collection will penetrate people's lives even more deeply and introduce whole new sets of linkable and identifiable private data. During this explosion of data collection technology, human involvement also changes qualitatively: where smart phones and social networks require a significant amount of active participation and awareness, humans will be mostly passive and unaware of data collection by the foreseen flood of smart things. However, it is difficult to give reliable estimates of the extent of data collection and the degree of penetration into the everyday of people's lives as forecasts greatly differ [44, 43, 46].

*Thing interaction:* With increasing numbers of smart connected things, it will become important what interfaces they provide to humans for configuration, debugging and interaction. We have witnessed a development from RFID tags with little interaction capabilities to sensors and devices featuring limited interfaces, such as LEDs and small displays responding to buttons and touches. Predictions that foresee mainly more haptic interfaces [2] and the smart phone as central mediator between humans and things based on web interfaces [3] seem very certain, as we are already starting to see these developments today. It is an interesting open question, whether web-based interfaces will prevail as primary interfaces to things. Other prediction such as Intel's for thought-driven control [47] by 2020 remain very dubious. The lack of interfaces and interaction mechanisms can pose privacy threats as observed in RFID [67]. On the other hand, too complex interfaces (as e.g. observed in social networks) do not seem to help either, while increasingly personalized interaction such as speech or, at the extreme, thoughts will even arise new privacy issues.

*System interaction:* Apart from the interaction with things, it is interesting how higher level interaction of humans with the IoT, i.e. with groups of things, systems and services, will develop. Similar arguments and predictions as for thing interaction apply also to system interaction. The main difference is that the whole environment is envisioned to serve for interaction with humans based on elaborate interfaces realized through the collaboration and coordination of many smart things and their specific capabilities. Less coherent predictions foresee thought-driven interaction and mechanisms that appeal to all human senses. An environment where humans are pervasively and passively exposed to interaction and feedback from IoT systems must be carefully designed not to violate their privacy.

*Lifecycle:* Usually, a very simple lifecycle of a thing is assumed. It is sold, used by its owner and finally disposed of and no information is stored by the thing in this process. The protest against the privacy-violating Benetton campaign [6] in 2003 have proven that this model is oversimplying: Benetton planned to equip its entire product line with RFID tags, that would remain active after garments had been sold. This change of ownership in the lifecycle of the RFID tag was not considered. In the IoT, smart things are predicted to store extensive information about their own history throughout their entire lifecylcle [41]. Furthermore, many everyday things, such as light bulbs, are designed to be easily exchangeable as should be their smart equivalents. It remains to be seen, if lifecycles become even more dynamic, e.g. with business models evolving around borrowing instead of owning smart things such as medical equipment or tools. With increasing storage of data and transitions during the lifecycle of a thing, managing security and privacy aspects will become more difficult.

*From vertical to horizontal:* Today's IoT-like systems are mostly separate vertically-integrated solutions, often based on custom technology, protocols and architectures. Direct collaboration between two such systems is the exception. With increasing standardization of protocols and platforms, horizontal integration of systems, i.e. increasingly local and distributed collaboration, is predicted. E.g. smart meters could cooperate with other meters in the neighborhood and directly switch on and off the household's appliances instead of only pushing consumption data to the central utility provider. Despite the promising advantages of horizontal integration performance- and functionality-wise, central solutions still seem to prevail and it remains unclear if and when the potential of horizontal integration can be realized. The envisioned collaboration of systems with different purposes and manufacturers can, however, induce privacy threats and security breaches. While, on the other hand, the locality of information flows in horizontally integrated systems is inherently more privacy-preserving than in their vertically-integrated, centralized counterparts.

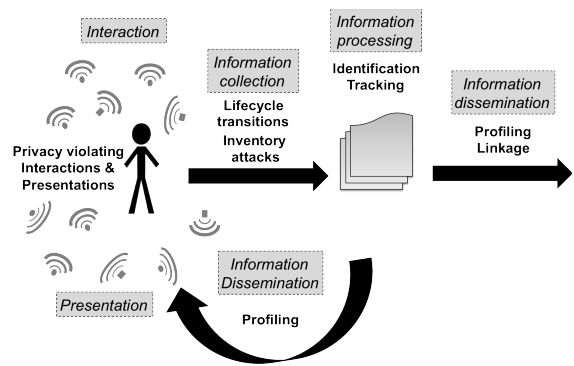

**Figure 2.** Threats in the Reference Model.

## 4. PRIVACY THREATS AND CHALLENGES IN THE IOT

The evolving nature of the IoT regarding technologies and features as well as the emerging new ways of interaction with the IoT lead to specific privacy threats and challenges. This section presents our classification of those threats. Figure 2 arranges them into the five different phases from our reference model (Section 2.2) according to where they are most prone to appear.

Each of the seven threat categories is analyzed in four-steps: First, a definition and characterization of the threat category with concrete instances of privacy violations is given. Second, we analyze how the IoT evolution impacts, changes and aggravates the particular threat. Here, Table II serves as a summary of two or three selected features with most impact on a particular threat category. Third, we identify approaches and counter-measures from related work and ask whether they be applied also in the IoT or are insufficient. Fourth, we present the main technical challenges and potential approaches to overcome those threats in the IoT. Of course, the threat categories are not totally disjunct and we make sure to point out dependencies and overlaps between different threats. Note that our perspective and classification thereby focusses on privacy threatening functionality in the IoT to allow better reasoning over the technical roots of threats and possible counter-measures than a user- or incident-centric perspective, as e.g. taken by [68, 15].

### 4.1. Identification

Identification denotes the threat of associating a (persistent) identifier, e.g. a name and address or a pseudonym of any kind, with an individual and data about him. The threat thus lies in associating an identity to a specific privacy violating context and it also enables and aggravates other threats, e.g. profiling and tracking of individuals or combination of different data sources.

The threat of identification is currently most dominant in the information processing phase at the backend services of our reference model, where huge amounts of information is concentrated in a central place outside of the subject's control. In the IoT also the interaction and collection phase will become relevant, as the impact of the evolving technologies and interconnection and interaction features aggravates the threat of identification, as identified in Table II: First, *surveillance camera technology* is increasingly integrated and used in non-security contexts, e.g. for analytics and marketing [69, 70]. As facial databases (e.g. from Facebook) become available also to non-governmental parties like marketing platforms [71], automatic identification of individuals from camera images is already a reality. Second, the increasing (wireless) interconnection and vertical communication of everyday things, opens up possibilities for identification of devices through *fingerprinting*. It was recognized already for RFID technology that individuals can be identified by the aura of their things [29]. Third, *speech recognition* is widely used in mobile applications and huge data-bases of speech samples are already being built. Those could potentially be used to recognize and identify individuals, e.g. by governments requesting access to that data [72]. With speech recognition evolving as a powerful way of interaction with IoT systems (Table I) and the proliferation of cloud computing for processing tasks, this will further amplify the attack vector and privacy risks.

Identity protection and, complementary, protection against identification is a predominant topic in RFID privacy (Section 3.1), but has also gained much attention in the areas of data anonymization [73, 74, 75], and privacy-enhancing identity management [76, 77]. Those approaches are difficult to fit to the IoT: Most data anonymization techniques can be broken using auxiliary data [78, 79, 80], that is likely to become available at some point during the IoT evolution. Identity management solutions, besides relying heavily on expensive crypto-operations, are mostly designed for very confined environments, like enterprise or home networks and thus difficult to fit to the distributed, diverse and heterogeneous environment of the IoT. Approaches from RFID privacy due to similarities in resource constraints and numbers of things are the most promising. However, those approaches do not account for the diverse data sources available in the IoT as e.g. camera images and speech samples.

We see as a main challenge the design of IoT systems that favor local over centralized processing, horizontal over vertical communications, such that less identifying data is available outside the personal sphere of a user and attack vectors for identification are reduced. Since identification is not always possible or even desired to prevent, it is important that users are made aware according to our privacy definition (Section 2.1), which in itself is a major challenge.

### 4.2. Localization and Tracking

Localization and tracking is the threat of determining and recording a person's location through time and space. Tracking requires *identification* of some kind to bind continuos localizations to one individual. Already today, tracking is possible through different means, e.g. GPS, internet traffic, or cell phone location. Many concrete privacy violations have been identified related to this threat, e.g. GPS stalking [81], disclosure of private information such as an illness [82], or generally the uneasy feeling of being watched [83]. However, localization and tracking of individuals is also an important functionality in many IoT systems. These examples show that users perceive it as a violation when they don't have control over their location information, are unaware of its disclosure, or if information is used and combined in an inappropriate context. This coincides with our definition of privacy (Section 2.1).

In the immediate physical proximity, localization and tracking usually do not lead to privacy violations, as e.g. anyone in the immediate surrounding can directly observe the subject's location. Traditionally, localization and tracking thus appears as a threat mainly in the phase of information processing, when locations traces are built at backends outside the subject's control. However, the IoT evolution will change and aggravate this threat in three ways: First, we observe an increasing use of location-based services (LBS). IoT technologies will not only support the development of such LBS and improve their accuracy but also expand those services to indoor environments e.g. for smart retail [84, 85]. Second, as data collection becomes more passive, more pervasive and less intrusive, users become less aware of when they are being tracked and the involved risks. Third, the increasing interaction with smart things and systems leaves data trails that not only put the user at risk of identification but also allow to track his location and activity, e.g. swiping an NFC-enabled smart phone to get a bus ticket or using the cities' smart parking system. With these developments, the threat of localization and tracking will appear also in the interaction phase, making the subject trackable in situations where he might falsely perceive physical separation from others, e.g. by walls or shelves, as privacy.

Research on location privacy has proposed many approaches that can be categorized by their architectural perspective [86] into (i) client-server, (ii) trusted third party, and (iii) distributed/peer-to-peer. However, those approaches have been mostly tailored to outdoor scenarios where the user actively uses a LBS through his smart phone. Thus, these approaches do not fit without significant modifications to the changes brought by IoT. The main challenges we identify are (i) awareness of tracking in the face of passive data collection, (ii) control of shared location data in indoor environments, and (iii) privacy-preserving protocols for interaction with IoT systems.

**Table II.** Summary of the impact of the evolving features on the seven categories of privacy threats. Statements in italics indicate that impact is possibly ambiguous.

| | Technology | Size | Interconnection | Data collection | Thing Interaction | System Interaction | Lifecycle | Vertical vs. horizontal |
|---|---|---|---|---|---|---|---|---|
| **Identification** | Cameras, face recognition | | Fingerprinting | | | Speech, cloud interfaces | | |
| **Tracking** | Indoor LBS | | | Decreasing awareness | | Data trails | | |
| **Profiling** | | Explosion of data sources | | Qualitatively new sets of data | | | | |
| **Interaction & Presentation** | | | | | *Presentation media* | Pervasive interaction with users | | |
| **Lifecycle transitions** | | | | Product history log | | | Exchangeability | Sensitive data on devices |
| **Inventory attacks** | *Diversification* | | Wireless communication | | | | | |
| **Linkage** | | | | Decreasing transparency | | | | *Drives linkage locally* |

### 4.3. Profiling

Profiling denotes the threat of compiling information dossiers about individuals in order to infer interests by correlation with other profiles and data. Profiling methods are mostly used for personalization in e-commerce (e.g. in recommender systems, newsletters and advertisements) but also for internal optimization based on customer demographics and interests. Examples where profiling is leads to a violation of privacy violation are price discrimination [87, 88], unsolicited advertisements [83], social engineering [89], or erroneous automatic decisions [48], e.g. by facebooks automatic detection of sexual offenders [90]. Also, collecting and selling profiles about people as practiced by several data market places today is commonly perceived as a privacy violation. The examples show that the profiling threat appears mainly in the dissemination phase, towards third parties, but also towards the subject itself in form of erroneous or discriminating decisions.

The impact of the evolving features in the IoT is mainly twofold, as shown in Table II: First, the IoT evolution leads to an explosion of data sources as more and more everyday things get connected. Second, while data collection thus increases quantitatively by orders of magnitude, it also changes qualitatively as data is collected from previously inaccessible parts of people's private lives. Additionally, the aggravation of identification and tracking threats further fuels the possibilities for profiling and danger from dubious data-selling businesses.

Existing approaches to preserve privacy include client-side personalization, data perturbation, obfuscation and anonymization, distribution and working on encrypted data [91, 48]. These approaches can possibly be applied to IoT scenarios but must be adapted from the usual model that assumes a central database and account for the many distributed data sources which are expected in the IoT. This will require considerable efforts for recalibration of metrics and redesign of algorithms, as e.g. recent work in differential privacy for distributed data sources shows [92]. After all, data collection is one of the central promises of the IoT and a main driver for its realization. We thus see the biggest challenge in balancing the interests of businesses for profiling and data analysis with users' privacy requirements.

### 4.4. Privacy-violating interaction and presentation

This threat refers to conveying private information through a public medium and in the process disclosing it to an unwanted audience. It can be loosely sketched as shoulder-surfing but in real-world environments.

Many IoT applications, e.g. smart retail, transportation, and healthcare, envision and require heavy interaction with the user. In such systems, it is imaginable that information will be provided to users using smart things in their environment (Table I), e.g. through advanced lighting installations, speakers or video screens. Vice versa, users will control systems in new intuitive ways using the things surrounding them, e.g. moving, touching and speaking to smart things. However, many of those interaction and presentation mechanisms are inherently public, i.e. people in the vicinity can observe them. This becomes a threat to privacy when private information is exchanged between the system and its user. In smart cities e.g. a person might ask for the way to a specific health clinic. Such a query should not be answered e.g. by displaying the way on a public display nearby, visible to any passers-by. Another example are recommendations in stores that reflect private interests, such as specific diet food and medicine, movies or books on precarious topics. Due to its close connection to interaction and presentation mechanisms, the threat of privacy-violating interactions and presentation appears primarily in the homonymous phases of our reference model.

Since such advanced IoT services are still in the future, privacy violating interactions have not received much attention from research. Interaction mechanisms are however crucial to usable IoT systems and privacy threats must consequently be addressed. We identify two specific challenges that will have to be solved: First, we need means for automatic *detection of privacy-sensitive content*. It is easily imaginable that the provisioning of content and rendering it for the user are handled in two steps by two different systems: E.g. company *A* generates recommendations for customers of a store, which are then delivered to the customer by company *B's* system: either by special lighting and the use of speakers or through a push to his smart phone. How to choose between those two interaction mechanisms, one public one private? Should company *A* mark privacy sensitive content or should company *B* detect it? How can company *B* (committed to privacy) protect itself from *A's* lax privacy attitude? Automatic detection of privacy-sensitive content can help to decide these questions. Second, with the previous point in mind, *scoping* will be necessary, i.e. how can we scope public presentation medium to a specific subgroup of recipients or a specific physical area? This approach would proof useful to support users, which have no smart phone (or any other device providing a private channel for interactions and presentations). However, it will be difficult to accurately determine the captive audience of a particular presentation medium, separate the intended target group and adjust the scope accordingly. E.g. what if the target user is in the midst of a group of people?

Applications for privacy-preserving pervasive interaction mechanisms are, e.g. smart stores and malls, smart cities and healthcare applications. Here, it would certainly be an achievement to provide similar levels of privacy as people would expect in the contexts of their everyday conversations, i.e. interactions with their peers.

### 4.5. Lifecycle transitions

Privacy is threatened when smart things disclose private information during changes of control spheres in their lifecycle. The problem has been observed directly with regard to compromising *photos and videos* that are often found on used cameras or smart phones – in some cases, disturbing data has even been found on 'new' devices [93]. Since privacy violations from lifecycle transitions are mainly due to the collected and stored information, this threat relates to the information collection phase of our reference model.

Two developments in the IoT will likely aggravate issues due to the lifecycle of things (Table II). First, smart things will interact with a number of persons, other things, systems, or services and amass this information in product history logs. In some applications, such data is highly sensitive, e.g. health-data collected by medical devices for home-care. But also the collection of simple usage-data (e.g. location, duration, frequency) could disclose much about the lifestyle of people. Already today, detailed usage logs are maintained for warranty cases in TV sets, notebooks or cars. Second, as exchangeable everyday things such as lightbulbs become smart, the sheer numbers of such things entering and leaving the personal sphere will make it increasingly difficult to prevent disclosure of such information.

Despite obvious problems with the lifecycle of todays smart phones, cameras, and other storage devices this threat has not been properly addressed. The lifecycle of most consumer products is still modeled as buy-once-own-forever and solutions have not evolved beyond a total memory wipe (e.g. before selling a phone) or physical destruction (e.g. before disposal of a hard drive). Smart things could, however, feature a much more dynamic lifecycle, with things being borrowed, exchanged, added and disposed freely. We thus identify the requirement for flexible solutions that will undoubtedly pose some challenges: *Automatic detection of lifecycle transitions* of a smart thing will be required to implement convenient privacy lifecycle management mechanisms. E.g. a smart rubbish bin could automatically cleanse all items in it from private information, such as medicine prescriptions on a smart pill box. It will be difficult, though, to automatically distinguish between different lifecycle transitions as e.g. lending, selling or disposing of an item and taking the appropriate action. Certain lifecycle transitions, e.g. borrowing a smart thing, will require *locking private information temporarily*, e.g. the readings of a vital signs monitor. Once the device has returned to its original owner, the private data can be unlocked and the original owner can continue to use it seamlessly.

### 4.6. Inventory attack

Inventory attacks refer to the unauthorized collection of information about the existence and characteristics of personal things.

One evolving feature of the IoT is interconnection (Table 3.2). With the realization of the All-IP and end-to-end vision, smart things become query-able over the Internet. While things can then be queried from anywhere by legitimate entities (e.g. the owner and authorized users of the system), non-legitimate parties can query and exploit this to compile an inventory list of things at a specific place, e.g. of a household, office building, or factory. Even if smart things could distinguish legitimate from illegitimate queries, a fingerprint of their communication speeds, reaction times and other unique characteristics could potentially be used to determine their type and model. With the predicted proliferation of wireless communication technology, fingerprinting attacks could also be mounted passively, e.g. by an eavesdropper in the vicinity of the victim's house. Since inventory attacks are mainly enabled by the increasing communication capabilities of things, the threat arises in the information collection phase of our reference model.

The impact of new technologies on this threat is not yet clear. On the one hand, we expect the diversification of technologies in the IoT as more and more different things become smart. Diversification increases the attack vector for fingerprinting, as e.g. observed with the many diverse configurations of web browsers [94]. On the other hand, at some point in time we expect the establishment of certain standards for communication and interaction that could reduce such differences.

Manifold concrete privacy violations based on inventory attacks are imaginable or have actually happened. First, *burglars* can use inventory information for targeted break-ins at private homes, offices and factories, similar to how they already use social media today to stake out potential victims [95]. Note that a comprehensive inventory attack could then also be used to profile the anti-burglar system down to every last presence sensor. Second, *law enforcement* and other authorities could use the attack to conduct (unwarranted) searches. Third, *private information* is disclosed by the possession of specific things, such as personal interests (e.g. books, movies, music) or health (e.g. medicine, medical devices). Fourth, efforts for *industrial espionage* can be complemented through an inventory attack, as noted by Mattern [3].

Radomirovic [13] and Van Deursen [29] have recognized the danger of profiling through fingerprinting in the context of RFID. However, with RFID the problem is at a much more local scope as RFID tags can be read only from a close distance and queries are mostly restricted to reading the tag's identifier. As analyzed above, the problem will aggravate in the IoT evolution as the attack vector is greatly increased by increasing proliferation of wireless communications, end-to-end connectivity, and more sophisticated queries. In order to thwart inventory attacks in the IoT, we identify the following two technical challenges: First, smart things must be able to *authenticate* queries and only answer to those by legitimate parties to thwart active inventory attacks through querying. Research in lightweight security provides useful approaches for authentication in resource-constrained environments. Second, mechanisms that ensure *robustness against fingerprinting* will be required to prevent passive inventory attacks based on the communication fingerprint of a smart thing. Inventory attacks will undoubtedly be difficult to counter. The fact that the use of PETs, though meant to protect privacy, can actually make fingerprinting even easier, leaves hiding in the (privacy-ignorant) masses currently as the most viable but clearly suboptimal solution

[94]. However, an IoT system that discloses comprehensive information about its owner's possessions is not likely to gain acceptance.

### 4.7. Linkage

This threat consists in linking different previously separated systems such that the combination of data sources reveals (truthful or erroneous) information that the subject did not disclose to the previously isolated sources and, most importantly, also did not want to reveal. Users fear *poor judgement* and *loss of context* when data that was gathered from different parties under different contexts and permissions is combined [15, 48]. Privacy violations can also arise from *bypassing privacy protection mechanisms*, as the risks of unauthorized access and leaks of private information increases when systems collaborate to combine data sources. A third example of privacy violations through linkage of data sources and systems is the increased risk of *re-identification of anonymized data*. A common approach towards protecting privacy is working on anonymized data only, but the act of combining different sets of anonymous data can often enable re-identification through unforeseen effects [78, 79, 80]. The examples show that the threat of linkage primarily appears in the information dissemination phase (Figure 2).

The threat of linkage will aggravate in the IoT evolution for two main reasons. First, horizontal integration will eventually link systems from different companies and manufacturers to form a heterogeneous distributed system-of-systems delivering new services that no single system could provide on its own. Successful collaboration will above all require an agile exchange of data and controls between the different parties. However, as horizontal integration features more local data flows than vertical integration it could provide a way to enhance privacy. Second, the linkage of systems will render data collection in the IoT even less transparent than what already is expected from the predicted passive and unintrusive data collection by smart things.

Threats from linking different systems and information sources are not entirely new. They can already be observed in the domain of online social networks (OSN) and their applications. However, this involves only two parties (i.e. the OSN and the third party application), while the IoT is expected to feature services that depend on the interaction and collaboration of many coequal systems. Here, we identify three technical challenges for privacy-enhanced systems-of-systems: First, *transparency* about what information a system-of-systems shares with whom is crucial to gain user acceptance. Second, *permission models and access control* must be adapted to the plurality of stakeholders collaborating in linked systems. Third, *data anonymization techniques* must work on linked systems and be robust against combination of many different sets of data. E.g. it will be interesting how concepts like differential privacy can be fitted to such multi-stakeholder multi-systems scenarios.

## 5. CONCLUSION

This paper motivates the need for a detailed analysis of privacy threats and challenges in the Internet of Things. We dissect this complex topic into a four-step approach: First, we provide a formal basis for discussing privacy in the IoT by concisely framing our notion of privacy and the applied reference model. A short review of relevant privacy legislation identifies clear insufficiencies and further motivates the need for a detailed assessment of privacy threats. In the second step, we acknowledge that the Internet of Things is constantly evolving and cannot be reduced to the sum of the technologies it builds upon. Here, our discussions of evolving technologies and features provide both a general and privacy-focused view on the past, present and future evolution of the IoT. Thirdly, we summarize existing privacy threats into seven categories and review them in the light of the evolving IoT. *Identification*, *tracking* and *profiling* are long known threats that, as we show, will be greatly aggravated in the IoT. The four threats of *privacy-violating interactions and presentations*, *lifecycle transitions*, *inventory attacks* and *information linkage* arise later in the IoT evolution. They represent partly new threats that have only been scratched in the related work, but can become very dangerous with regard to the predicted evolution of the IoT. The arrangement of threats in our reference model provides a clear idea of where threats appear and where to approach them conceptually. Finally, technical challenges are discussed in the context of each threat that provide clear directions for future research.

We consider that profiling remains one of the most severe threats: Our analysis shows that it is greatly aggravated and that other threats like identification or tracking, though each provoking different very specific privacy violations, add to its dangers by supplying even more linkable data. At the same time, business models that depend heavily on profiling have enjoyed tremendous success and so the trend for *big data* continues, fueled by the IoT's central promise for fine-grained and ubiquitous data collection. Here, the challenge consists in designing privacy-aware solutions for the IoT that allow to balance business interests and customers' privacy requirements.

We consider privacy-violations in the interaction and presentation phase an important future threat, because of the corresponding interaction mechanisms with smart things and systems that are just evolving and are rather unique to the IoT. The involved technical challenges have hence received little attention in the related work so far and require new ways of using technology as well as a fair amount of foresight and sensitivity for privacy implications. We aim to design technical solutions that allow users to interact with IoT systems in manifold ways, while affording privacy protection similar to the intuitive understanding of privacy people apply in real world situations.

Finally, we stress two core thoughts, that our work suggests for a privacy-aware Internet of Things: First, the IoT is evolving – privacy is a constant challenge and must be faced with the necessary foresight. Second, a fruitful outcome requires coordinated action to provide technical solutions supported by the corresponding legal framework.